\documentclass[aps,prb,twocolumn,superscriptaddress]{revtex4-2}

\usepackage{amsmath,amssymb}
\usepackage{graphicx}
\usepackage{epstopdf}
\usepackage{bm}
\usepackage{xcolor}

\newcommand{\CTO}     {Cu$_3$TeO$_6$}
\newcommand{\te}	{$^{125}$Te}
\newcommand{\slr} 	{$T_1^{-1}$}

\newcommand{\slrt} 	{$(T_1T)^{-1}$}

\newcommand{\kk} 	{$\mathcal{K}$}

\newcommand{\TN} 	{$T_\text{N}$}
\newcommand{\TS} 	{$T_\text{S}$}

\begin{document}

\title{Unusual spin pseudogap behavior in the spin web lattice Cu$_3$TeO$_6$ probed by \te\ nuclear magnetic resonance}

\author{Seung-Ho Baek}
\email[]{sbaek.fu@gmail.com}
\affiliation{Department of Physics, Changwon National University, Changwon 51139, Korea}
\affiliation{Department of Materials Convergence and System Engineering, Changwon National University, Changwon 51139, Korea}
\author{Hyeon Woo Yeo}
\affiliation{Department of Physics, Changwon National University, Changwon 51139, Korea}
\author{Jena Park}
\affiliation{Department of Physics, Chung-Ang University,  Seoul 156-756,
Korea}
\author{Kwang-Yong Choi}
\affiliation{Department of Physics, Sungkyunkwan University, Suwon 16419, Republic of Korea}
\author{Bernd B\"uchner}
\affiliation{Institut f\"ur Festk\"orper- und Materialphysik and W\"urzburg-Dresden Cluster of Excellence ct.qmat, Technische Universit\"at Dresden, 01062 Dresden, Germany}
\affiliation{IFW Dresden, Helmholtzstr. 20, 01069 Dresden, Germany}

\date{\today}


\begin{abstract}

We present a \te\ nuclear magnetic resonance (NMR) study in the three-dimensional spin web lattice \CTO\, which
harbors topological magnons. The \te\ NMR spectra and the Knight shift \kk\ as a function of temperature show a drastic change at $T_\text{S}\sim 40$ K much lower than the N\'eel ordering temperature $T_\text{N}\sim 61$ K,  providing evidence for the first-order structural phase transition within the magnetically ordered state.
Most remarkably, the temperature dependence of the spin-lattice relaxation rate \slr\ unravels spin-gap-like magnetic excitations, which sharply sets in at $T^*\sim 75$ K, the temperature well above $T_\text{N}$. The spin gap behavior may be understood by weakly dispersive optical magnon branches of high-energy spin excitations originating from the unique corner-sharing Cu hexagon spin-1/2 network with low coordination number.

\end{abstract}

\maketitle


\section{Introduction}

In the past decades, research of quantum spin systems has mainly focused on the search for quantum states of matter, such as spin singlets (dimers), valence bond solids \cite{read89}, and quantum spin liquids \cite{balents10} in which strong quantum fluctuations depending on dimensionality and geometry of the underlying lattice on which the spins reside prevent conventional spin ordering.
Recently, however, the observation of topological band structures in fermionic systems \cite{hasan10a} has also raised the fundamental interest in magnetically ordered systems with magnon excitations due to the possibility of hosting topologically nontrivial Dirac and Weyl magnons.
Indeed, the topological magnons could be realized in various quantum magnets on frustrated lattices such as kagome \cite{chisnell15}, honeycomb \cite{owerre16,fransson16}, and pyrochlore \cite{onose10,zhang13,li16,mook16} geometries. As can be inferred from their nonbipartite lattice structures, topological magnon materials possess optical magnon bands in addition to acoustic magnon branches, which lead to topological linear band crossings in spin-wave spectra.  However, little is known about how the two distinct magnon bands affect static and dynamic magnetisms.

The Te-based cuprate \CTO\ is a three-dimensional (3D) antiferromagnetic (AFM) spin web lattice in a cubic structure with space-group $Ia\bar{3}$. Each copper ion is surrounded by an irregular octahedral arrangement of oxygen \cite{falck78}. It is shown that the nearest-neighbor exchange-coupling $J_1$ is dominant and forms the interesting 3D magnetic lattice structure consisting of eight almost planar Cu$^{2+}$ hexagons \cite{herak05,herak11,wang19}.  Each hexagon contains the Te ion at the center and shares vertices with six non-coplanar hexagons. All the hexagon planes are perpendicular to one of the four space diagonals of the cubic unit cell [see Fig. 1 (a)], forming four different orientations of hexagons. Consequently, the number of the first nearest-neighbors for each spin is only four, smaller than six or more in usual 3D spin lattices.

\CTO\ is a band gap insulator and undergoes nearly collinear AFM order along the [111] direction at $T_\text{N}\sim 61$ K with the Curie-Weiss (CW) temperature of $\Theta\sim -184$ K \cite{herak05,mansson12,he14}.
Interestingly, the bulk magnetic susceptibility $\chi(T)$ deviates from the CW behavior at $\sim 180$ K significantly higher than \TN, and it forms a maximum in the paramagnetic state above \TN\, suggesting the presence of short-ranged spin correlations.  Also, the spin ordering in a cubic structure, which is stabilized by the spin-orbit coupling \cite{chakraborty19}, seems to cause structural instability driven by the magneto-elastic effect, deep in the AFM state \cite{caimi06,choi08a}.
Furthermore, the recent theory of a Heisenberg spin model for this compound \cite{li17a} predicted the presence of topological Dirac and nodal line magnons, which was experimentally confirmed by inelastic neutron scattering \cite{yao18,bao18,bao20}.

Motivated by the unconventional magnetic and topological properties of \CTO, we carried out \te\ NMR measurements in a single crystal of \CTO\ to investigate the low-energy spin dynamics, local spin susceptibility, and the nature of structural distortion in the ordered state. {Although we are unable to assess the topological characters of magnon excitations because NMR is not a momentum-resolved probe, our NMR study unravels the significance of optical magnon bands in spin dynamics.}
The temperature dependence of the spin-lattice relaxation rate \slr\ demonstrates the presence of a large spin pseudogap in the magnetic excitation spectrum, which opens up at the well-defined temperature, $T^*=75$ K prior to the AFM ordering at $T_\text{N}=61$ K. We discuss a plausible scenario for the origin of the spin gap behavior --- a depletion of the low-energy spectral weight by the formation of high-energy optical spin excitations inherent to a 3D spin web lattice. Also, our NMR data demonstrate that \CTO\ undergoes
the structural phase transition at $T_S\sim 40$ K in the AFM ordered state.

\begin{figure*}
\centering
\includegraphics[width=0.8\linewidth]{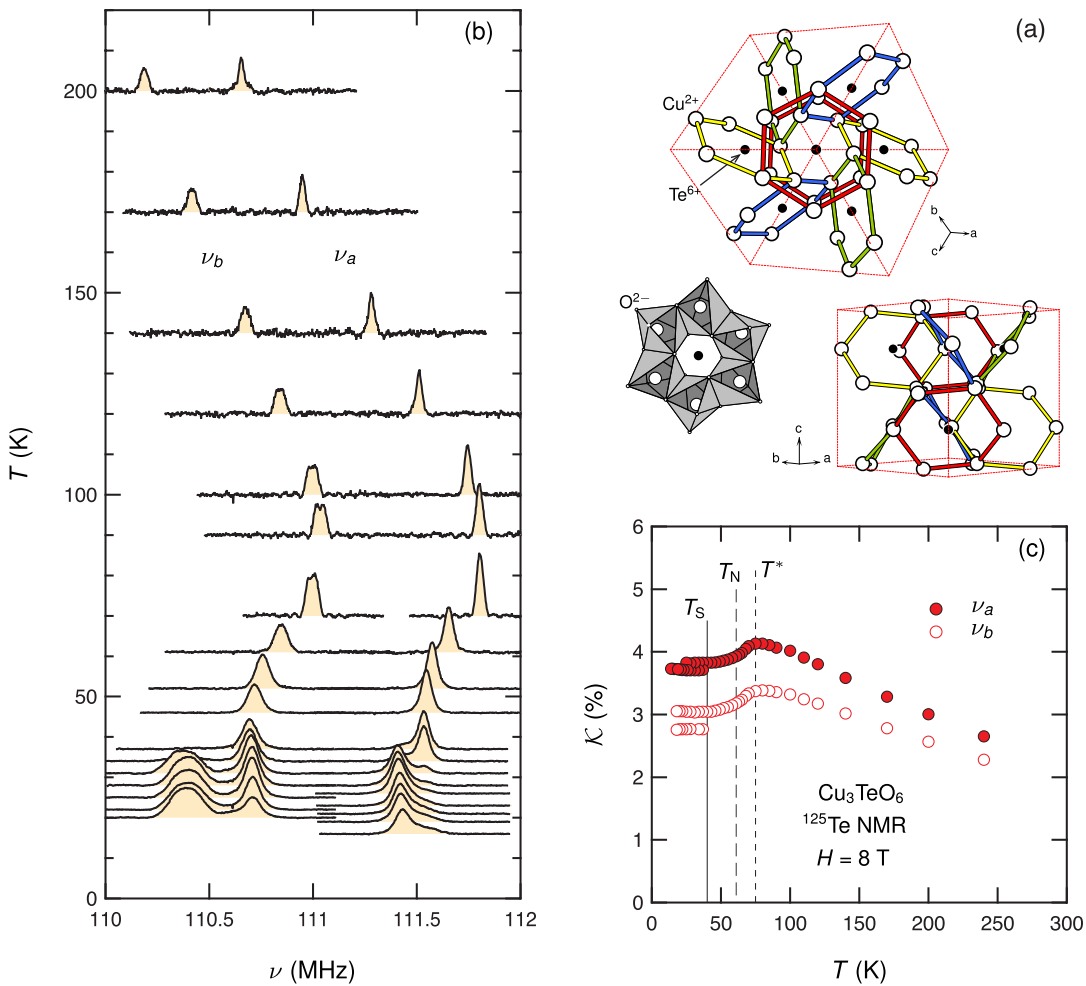}
\caption{(a) Cu$^{2+}$ magnetic lattice of \CTO\ taking into account only the nearest-neighbor exchange $J_1$, as viewed along the [111] direction (upper). There are eight Cu hexagons, each of which is formed by edge-sharing CuO$_6$ irregular octahedra (lower left). As $H$ is applied along the [110] direction (lower right), there exist two groups of inequivalent \te\ sites.
(b) Temperature dependence of \te\ NMR spectra measured at $8$~T applied along the [110] direction of \CTO. {The left axis refers to the temperature at which each spectrum was obtained, whereas the amplitude of the spectra is plotted in arbitrary units.} The Boltzmann correction was made by multiplying the temperature for each spectrum. The two NMR lines at $\nu_a$ and $\nu_b$ represent two inequivalent \te\ sites in this field direction. Below $T_\text{S}\sim 40$ K, the line at $\nu_a$ is replaced by a new line at a lower frequency, whereas there appears a new line at 110.4 MHz with a slight decrease of the line at $\nu_b$. 
(c) Knight shift \kk\ as a function of temperature. The local spin anisotropy is clearly detected. One could identify $T^*$ at the maximum and \TS\ from the abrupt changes of the lines. Since \kk\ is barely affected by the AFM transition,  \TN\ was determined by the derivative of the bulk susceptibility $\chi(T)$.
}
\label{spec}
\end{figure*}

\section{Experimental details}
\CTO\ single crystals were synthesized by using the TeO$_2$ self-flux method. A mixture of high purity chemical reagents of CuO and TeO$_2$ in a molar ratio of 1:1.3 was grounded for 40 min in mortar. The pelletized mixture was put into a covered alumina crucible and was placed in a box furnace. The furnace was heated up to 800$^{\circ}$ C and kept for 24 h and then slowly decreased to 600$^{\circ}$ C for 100 h. After cooling to room temperature, shiny single crystals \CTO\ was obtained from the mechanically removed flux.
The size of the single crystal used in this NMR study is $1.2\times 2\times 4$ mm$^3$.

\te\ (nuclear spin $I=1/2$) NMR measurements were performed in the temperature range of 10--300 K in the external magnetic field of 8 T.
Since \te\ is located at the center of the planar Cu hexagons in \CTO, the hyperfine couplings of \te\ with Cu spins should be anisotropic depending on the field direction with respect to each planar hexagon. Accordingly, in an arbitrary field direction, one could expect up to four different \te\ lines, making the analysis of the obtained data extremely complicated. So we decided to apply the external field $H$ along the [110] direction in order to form only two, instead of four, inequivalent \te\ sites, each of which belongs to the two hexagon pairs that are either $45^\circ$ tilted or parallel with respect to $H$, as depicted in the lower panel of Fig. 1 (a).

The \te\ NMR spectra were acquired by a standard spin-echo technique with a typical $\pi$/2 pulse length of $\sim 2$ $\mu$s. At low temperatures below $T_\text{S}=40$ K, we obtained the NMR spectra by a frequency-sweep method.   The nuclear spin-lattice relaxation rate \slr\ was measured by a saturation method, and determined by fitting the recovery curve of the nuclear magnetization $M(t)$ to a single exponential function $\exp(-t/T_1)$.


\section{Results}
\subsection{\te\ NMR spectra and the Knight shift}
Figure 1 (b) presents the temperature dependence of the \te\ NMR spectra measured at 8 T applied along the [110] direction.  We observed the two NMR lines which are well separated, as expected from the two different groups of \te\ sites as shown in the lower panel of Fig. 1(a). We will refer to the resonance frequencies of the two lines as $\nu_a$ and $\nu_b$ ($\nu_a>\nu_b$), respectively, for convenience. The complicated lines observed below 40 K are ascribed to a lowering of the crystal symmetry (see below).
Interestingly, the line at $\nu_b$ is notably broader than that at $\nu_a$ which remains sharp.
Note that the line at $\nu_b$ is resolved into two near 100 K, indicating a slight misalignment of the crystal in such a way as to mainly affect the line at $\nu_b$. This could be understood by the fact that it is relatively easier to make the four hexagon planes parallel to $H$ simultaneously, as compared to aligning other four hexagon planes with the same angle of $45^\circ$ with $H$. Therefore, we assign the line at $\nu_a$ to the \te\ sites at the hexagons whose plane is parallel to $H$, and the line at $\nu_b$ to the other sites at the hexagons whose plane makes an angle of $\sim 45^\circ$ with $H$.

\begin{figure*}
\centering
\includegraphics[width=\linewidth]{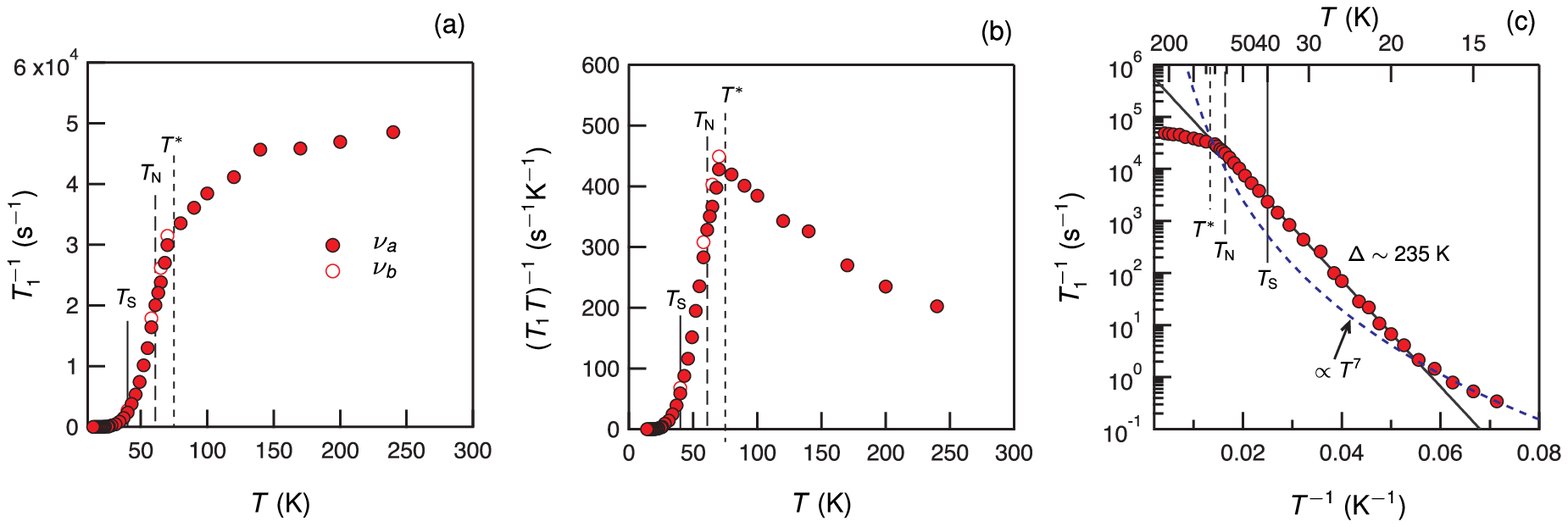}
\caption{(a) and (b) \te\ spin-lattice relaxation rate \slr\ and \slrt\ as a function of temperature, respectively, measured mainly at $\nu_a$ at 8 T. The sharp anomaly of \slrt\ is found at $T^*\sim 75$ K. Just below $T^*$, \slr\ decreases exponentially, with a slight change at $T_\text{N}\sim61$ K. (c) \slr\ as a function of inverse temperature $1/T$. \slr\ decreases more than 4 decades in the 40 K temperature range, revealing the activation behavior with the gap $\Delta\sim 235$ K that is equivalent to $\sim20$ meV. At low temperatures below 20 K, \slr\ deviates from the activation law, being described by the Raman relaxation rate, $T_1^{-1}\propto T^7$.
}
\label{t1t}
\end{figure*}

The temperature dependence of $\nu_a$ and $\nu_b$ is plotted in Fig. 1 (c) in terms of the Knight shift, which is defined by $\mathcal{K} = (\nu-\nu_0)/\nu_0\times 100$ \%, where $\nu_0$ is the unshifted Larmor frequency. It reveals the presence of a sizable {anisotropy of the local spin susceptibility and/or the hyperfine coupling}, despite the cubic symmetry of the crystal {which yields the nearly isotropic bulk magnetic susceptibility $\chi(T)$.} The anisotropy of the Knight shift gradually decreases with increasing temperature, seemingly collapsing at high temperatures. Except for the anisotropy, the temperature dependence of \kk\ is well consistent with that of $\chi(T)$ \cite{herak05, choi08a}.
\kk\ increases with decreasing temperature, reaching the maximum at a temperature of $T^*\sim75$ K notably higher than $T_N\sim61$ K  \cite{choi08a}. The notable discrepancy between $T^*$ and \TN\ could be ascribed to quantum fluctuations which stabilize the short-ranged spin ordering preempting the N\'eel state.
Surprisingly, as will be discussed below, it turns out that $T^*$ is also the well-defined onset temperature of a {spin-gap-like behavior}.

{By plotting \kk\ vs $\chi$ (not shown), one could obtain the hyperfine coupling constants $A_\text{hf} \sim 141$ and 110.8 kOe/$\mu_B$ for the lines at $\nu_a$ and $\nu_b$, respectively. Note that these values are just rough estimations because $\chi(T)$ is nearly isotropic in the cubic crystal symmetry.}

Upon further cooling, \kk\ is quickly saturated toward a finite value at low temperatures, without any signature for the AFM transition at \TN.
This behavior resembles $\chi(T)$ obtained at a large external field of 5 T \cite{choi08a} which is close to 8 T we used, indicating that the hyperfine field at \te\ is proportional to the static magnetization enhanced by $H$ in the AFM ordered state, and thus the \te\ Knight shift precisely reflects the local magnetization.
In contrast to $\chi(T)$, however, the \te\ NMR spectra and the Knight shift drastically change at low temperatures below $T_\text{S}\sim 40$ K.  The line at $\nu_a$ is rapidly transferred to a lower frequency, whereas the original line partially remains below \TS, implying the coexistence of the two different structural phases. This is the typical behavior of a first-order structural phase transition, suggesting that the AFM spin ordering may drive a structural transition via the magnetoelastic coupling \cite{caimi06,choi08a}. We find that the line at $\nu_b$ shows a similar, but somewhat distinct behavior compared to that at $\nu_a$. {The new line emerging at a lower frequency is notably broader and stronger than the original line which decreases slowly with decreasing temperature. The different behavior of the two lines below \TS\ may reflect a complex structural modification which may have a non-trivial effect on the magnetic structure.}

Previously, optical reflectivity \cite{caimi06} and Raman scattering \cite{choi08a} measurements suggested the development of a local lattice distortion below 50 K. Taken together, we conclude that the lattice instability with a weak phonon softening \cite{choi08a} onsets near 50 K before undergoing the structural phase transition at $T_\text{S}\sim 40$ K. {At the moment, however, the lowered crystal symmetry generated below \TS\ is unknown, deserving x-ray and neutron studies in the future.} 

\subsection{The spin-lattice relaxation rate \slr}

Having discussed the static properties of \CTO, thus far, we now present the low-energy spin dynamics probed by the spin-lattice relaxation rate \slr\ as a function of temperature, mainly measured at $\nu_a$ as shown in Fig. 2 (a). We find that there is no essential difference of the \slr\ value between the lines at $\nu_a$ and $\nu_b$.
At high temperatures, we find that $T_1$ is extremely short ($\sim 20$ $\mu$s) so that it is difficult to be measured near room temperature.
With lowering temperature, \slr\ is gradually decreased, and it is strongly suppressed  below $T^*$, becoming vanishingly small at low temperatures. The drastic change in the relaxation rate at $T^*$ is clearly revealed in the plot of \slrt\ vs. $T$ in Fig. 2 (b). \slrt, which is a measure of spin fluctuations at low energies \cite{moriya63}, is gradually increased with decreasing temperature being followed by an abrupt drop. The sharp peak of \slrt\ at $T^*$ is reminiscent of a phase-transition-like anomaly rather than a crossover to a short-range magnetic order, as observed, for example, at $T_c$ in some superconducting materials \cite{yoshida07,dahm07}.
Remarkably, \slr\ is almost not affected by the subsequent occurrence of the AFM and structural phase transitions. Only weakly visible change of \slrt\ was observed at \TN, but the structural transition at \TS\ has no effect on \slr\ at all. This indicates that the nuclear relaxation process in both normal and ordered states is governed by unconventional spin excitation modes which overwhelm the low-energy quasiparticle excitations associated with the phase transitions.

The rapid decrease of spin fluctuations below $T^*$ suggests a spin-gap behavior.
In order to verify that a spin gap exists, we plotted \slr\ vs inverse temperature $1/T$ on a semilogarithmic scale as shown in Fig. 2 (c). It clearly shows that \slr\ is excellently described by the activation law  $T_1^{-1} \propto \exp(-\Delta/T)$,  in nearly 5 decades only within the 40-K temperature range, yielding the activation gap $\Delta\sim 235$ K. The deviation of \slr\ from the spin gap behavior occurs below $\sim 20$ K, suggesting that the relaxation process by gapless magnons becomes dominant over the gapped spin excitations at very low temperatures. {Here we stress that the observed spin-gap behavior is at complete odds with the development of critical slowing down on approaching \TN\ as well as with the conventional magnon dynamics expected for antiferromagnetically ordered systems.}
We stress that the conventional relaxation mechanisms such as Raman and Orbach processes are solely due to acoustic magnons. Thus, in the presence of weakly dispersive optical magnons, the  conventional processes provide no longer dominant relaxation channels.

\section{Discussion}

It is remarkable to note that \kk\ or the local spin susceptibility is suppressed just below $T^*$, the onset temperature of the spin pseudogap, as shown in Fig. 1 (b). The nearly simultaneous onset of the spin gap behavior in the spin fluctuation spectrum and the static short-range AFM ordering could be a strong signature that {localized spin excitations begin to appear at $T^*$}.

%
%
%

Interestingly, we find that the extracted gap value of $\Delta=235$ K, which corresponds to 20 meV, is very close to the energy scale of the optical magnon branches observed by inelastic neutron scattering \cite{yao18,bao18}. The similar energy scales, unless it is accidental, may imply that the nuclear relaxation process is governed by the high-lying spin excitations that is also directly relevant to the optical magnon branch formed in the ordered state. Namely, the spectral weight of spin excitations may be significantly shifted to higher energies, being strongly bunched up at energies near and above $\sim20$ meV, even at high temperatures.  As thermal fluctuations are reduced with decreasing temperature, the depleted spectral weight at low energies below $\sim20$ meV could cause the spin gap-like behavior of \slr\ below a characteristic temperature.  In support of this interpretation, the dominance of high-lying spin excitations for the nuclear relaxation could account for the extremely short $T_1$ observed near room temperature, as well as for the negligible influence of the magnetic and structural phase transitions on \slr. {Our interpretation is also consistent with the inelastic neutron scattering data (see Fig. 2d in Ref. \cite{yao18}), which shows a clear depletion of the total magnetic spectral weight at the low-energy window.}

%

%
The origin of the spin gap may be related to the fact that the coordination number of each spin in \CTO\ is only 4. This may cause each Cu hexagon to behave as a spin cluster weakly coupled to others, and the clusterlike interaction may cause the strong upshift of the spectral weight \cite{baek12c} giving rise to the unusual spin gap behavior in the short-ranged ordered state.
In this case, the suppression of the spin susceptibility below $T^*$ could be understood in terms of short-ranged AFM spin correlations related to the competition between intra- and inter-hexagon couplings.
Within this scenario, the lacking spin gap behavior in the static spin susceptibility below $T^*$ could mean that the static magnetism is dictated by the development of long-range magnetic order. In addition,
the sharp $T^*$ anomaly of spin fluctuations [see Fig.~2(b)] reflects that the optical magnons start to
gain the notable spectral weight in the short-range ordered temperature scale of $T^*$. This is contrasted
by the fact that the acoustic magnon is stabilized for temperatures below $\sim (0.4-0.5)$~\TN\  in the $k\approx 0$ limit.

Since the acoustic magnon fluctuations develop around a specific wavevector $\mathbf{Q}$ in contrast to the nearly non-dispersive optical magnon branches, one may raise a possibility that the spin gap behavior arises from the cancellation of the AFM fluctuations at the \te\ related to the acoustic magnon branch, if the hyperfine form factor $A_\text{hf}(\mathbf{Q})$ is zero. However, we believe that this scenario is highly unlikely in this complex 3D magnet which involves numerous hyperfine coupling channels. Moreover, the sharp onset of the spin gap behavior observed in the paramagnetic state at $T^*$ seems incompatible with this scenario.

Lastly, we discuss the deviation of \slr\ from the gap behavior at low temperatures [see Fig. 2(c)]. {Although it may be ascribed to the presence of the extrinsic relaxation caused by impurities, we consider an intrinsic relaxation channel.}
In addition to the strongly gapped optical magnons, there should exist spin excitation modes at low energies which are linked to the gapless acoustic magnons in the ordered state. Although the acoustic magnons appear to be ineffective for the $T_1$ process near \TN, it may contribute to the nuclear relaxation at very low temperatures.
In the AFM ordered state of \CTO, it was proposed that magnons are strongly coupled with phonons \cite{bao20}. Such mixing of magnons and phonons could generate the low-lying hybridized elementary excitations so-called \textit{quasimagnons}, which gives a $T^7$ temperature dependence for the Raman relaxation rate for nuclei at very low temperatures \cite{pincus61,turov72}. Indeed, the low temperature deviation of \slr\ from the activation law is excellently accounted for by the power law for the quasimagnons [see the dashed lines in Fig. 2(c)]. 
%


\section{Summary}
We have investigated the 3D spin-1/2 lattice \CTO\ by means of \te\ NMR. The \te\ NMR spectra and the Knight shift as a function of temperature proved that the first-order structural phase transition takes place at $T_\text{S}\sim40$ K within the magnetically ordered state. 
 On the other hand, the temperature dependence of the relaxation rate \slr\ unraveled a large spin gap of 235 K in the magnetic excitation spectrum, which opens up at $T^*=75$ K in the normal state prior to the AFM ordering at $T_\text{N}\sim 61$ K. The gapped excitations dominate the nuclear relaxation down to low temperatures, overwhelming the low-energy quasiparticle excitations in the vicinity of the AFM and structural phase transitions.
The closeness of the spin gap value to the energy scale of the optical magnon branch in the ordered state led us to conjecture that the spin gap behavior may originate from the large spectral weight of gapped high-energy spin excitations intrinsic to the spin-1/2 hexagon spin topology.
%

%
Although the origin of the spin-gap-like behavior remains an open question, it should be related to unconventional spin dynamics arising from the complex exchange paths in three dimensions as well as from the strong spin-lattice coupling that manifests itself in the structural phase transition within the AFM ordered state and the quasimagnon branch.
Together with the experimental observation of the topological Dirac and nodal line magnons, the dominance of the high-lying spin excitations with a large spin gap behavior makes \CTO\ a unique three-dimensional quantum magnet, calling for further detailed theoretical and experimental investigations. {A future challenge is to single out kinematics and dynamics of topological magnons out of the optical magnons.}


\begin{acknowledgments}
This work was supported by the National Research Foundation of Korea (NRF) grant funded by the Korea government(MSIT) (Grants No. NRF-2020R1A2C1003817
and No. 2020R1A2C3012367).
\end{acknowledgments}


\bibliography{mybib}

\end{document}